\documentclass[11pt]{article}
\usepackage{graphicx}
\usepackage[margin=1.25in]{geometry}
\usepackage[usenames,dvipsnames]{color}
\usepackage{url}
\usepackage{subcaption}   
\usepackage{wrapfig}      
\usepackage{cite}         

\usepackage[colorlinks = true,
            linkcolor = blue,
            urlcolor  = blue,
            citecolor = blue,
            anchorcolor = blue]{hyperref}

\newcommand\pubnumber{ }
\newcommand\pubdate{\today}

\def\Title#1{\begin{center} {\LARGE #1 } \end{center}}
\def\Author#1{\begin{center}{ \sc #1} \end{center}}

\newcommand\pubblock{\rightline{\begin{tabular}{l} \pubnumber\\
         \pubdate \end{tabular}}}
\newenvironment{Abstract}{\begin{quotation} \begin{center}
                       ABSTRACT
     \end{center}\bigskip  }{\end{quotation}}

\def\beq{\begin{equation}}
\def\eeq#1{\label{#1}\end{equation}}
\def\eeqn{\end{equation}}

\newenvironment{Eqnarray}%
   {\arraycolsep 0.14em\begin{eqnarray}}{\end{eqnarray}}
\def\beqa{\begin{Eqnarray}}
\def\eeqa#1{\label{#1}\end{Eqnarray}}
\def\eeqan{\end{Eqnarray}}

\let\bar=\overbar

\def\lsim{\mathrel{\raise.3ex\hbox{$<$\kern-.75em\lower1ex\hbox{$\sim$}}}}
\def\gsim{\mathrel{\raise.3ex\hbox{$>$\kern-.75em\lower1ex\hbox{$\sim$}}}}

\def\del{\partial}
\def\Dslash{\not{\hbox{\kern-4pt $D$}}}
\def\dslash{\not{\hbox{\kern-2pt $\del$}}}
\def\pslash{\not{\hbox{\kern-2pt $p$}}}
\def\ETmiss{\not{\hbox{\kern-4pt $E$}}_T}

\def\Dlr{\mathrel{\raise1.5ex\hbox{$\leftrightarrow$\kern-1em\lower1.5ex\hbox{$D$}}}}

\def\MSB{{\bar{M \kern -2pt S}}}
\def\msb{{\bar{\scriptsize M \kern -1pt S}}}

\def\drb{{\bar{\scriptsize D \kern -1pt R}}}

\newcommand\snowmass{\begin{center}\rule[-0.2in]{\hsize}{0.01in}\\\rule{\hsize}{0.01in}\\
\vskip 0.1in Submitted to the  Proceedings of the US Community Study\\ 
on the Future of Particle Physics (Snowmass 2021)\\ 
\rule{\hsize}{0.01in}\\\rule[+0.2in]{\hsize}{0.01in} \end{center}}

\begin{document}

\pubblock

\Title{Readout for Calorimetry at Future Colliders:\\
A Snowmass 2021 White Paper}

\bigskip 

\Author{Timothy Andeen, University of Texas at Austin \\
    Julia Gonski, Columbia University \\
    James Hirschauer, Fermi National Laboratory \\
    James Hoff, Fermi National Laboratory \\ 
    Gabriel Matos, Columbia University \\
    John Parsons, Columbia University}

\medskip


\medskip

 \begin{Abstract}
\noindent Calorimeters will provide critical measurements at future collider detectors. As the traditional challenge of high dynamic range, high precision, and high readout rates for signal amplitudes is compounded by increasing granularity and precision timing the readout systems will  become increasingly complex. This white paper reviews the challenges and opportunities in calorimeter readout at future collider detectors. 
\end{Abstract}

\snowmass

\def\thefootnote{\fnsymbol{footnote}}
\setcounter{footnote}{0}

\section{Introduction}

Nearly all particle detectors foreseen at future colliders, from detectors at the High Luminosity Large Hadron Collider (HL-LHC) currently under construction, to potential future circular and linear colliders, include some aspect of particle calorimetry. The need is clear. Calorimeters are the primary detectors used to measure the energy of neutral particles produced in collisions, and also provide an important measurement of charged particle energies, in addition to measurements of particle position and timing. Calorimeters are often split between two sections, an electromagnetic calorimeter (ECAL) in the front, followed by a hadronic calorimeter (HCAL). The ECAL specializes in measuring the energy deposited by e.g. electrons or photons, while the HCAL specializes in measuring the energy deposited by e.g. protons or pions. 

There are two common detector strategies used in the construction of calorimeters. A \textit{sampling calorimeter} samples the energy of the shower of particles resulting from the interaction of the primary particle. Dense material (such as lead or tungsten) is sandwiched between the active medium (such as liquid argon, silicon, scintillating material). The active medium is read out, resulting in an estimate of the particle's energy. Alternatively, a \textit{total absorption calorimeter} has an active medium that is also a dense, absorbing material. Both designs are currently being utilized in particle detection experiments. 

In this paper, two innovations for more performant calorimetry will be discussed. In \textit{particle flow} algorithms, measurements of the energies from calorimeters are frequently combined with the measurements from tracking detectors. In this way, more detailed information can be extracted from the collisions. In \textit{dual readout} calorimeters (Section~\ref{subsec:CMS}), complementary information is collected about the electromagnetic and hadronic components of particles to better characterize the full shower. Further, high-granularity imaging calorimeters could provide a detailed reconstruction of the particle shower. Precision timing could allow for ``4-D'' reconstruction of events, potentially helping to mitigate pileup at future hadron detectors. Future electronic readout systems will be designed with these algorithms, capabilities, and requirements in mind. 

The conceptual challenges for the readout of future calorimeters are two-fold. First, calorimeters have always uniquely required high dynamic range, high precision, and high readout rates for signal amplitudes. Future calorimeters will be required to measure energies of particles from hundreds of MeV (minimum ionizing particles) to up to tens of TeV at the bunch crossing rate. Second, requirements for increasing granularity and precision timing compound these challenges. The readout of calorimeters has traditionally been facilitated through the design of application specific integrated circuits, or ASICs. These ASICs face the primary engineering challenges inherent in designing low noise, low power, radiation tolerant devices. There are a number of strategies utilized to meet these challenges. However, the high granularity calorimeters at future colliders will require novel designs. 

This paper is organized into three sections. In Section~\ref{sec:hllhc}, we review the upgrades to the calorimeter detectors of the ATLAS and CMS experiments for the HL-LHC. These projects are well-advanced, and provide a background of what is technologically achievable today. In Section~\ref{sec:future}, we discuss planned future detectors at potential electron-positron colliders (linear or circular) and circular hadron colliders. Finally, in Section~\ref{sec:innov} we discuss several promising technological opportunities that may be useful in achieving the stringent design requirements of these future detectors.

   

\section{HL-LHC Calorimeter Upgrades}
\label{sec:hllhc}

\subsection{ATLAS Calorimeter Upgrade}

\begin{figure}[htbp]
\centering
\includegraphics[width=0.9\textwidth]{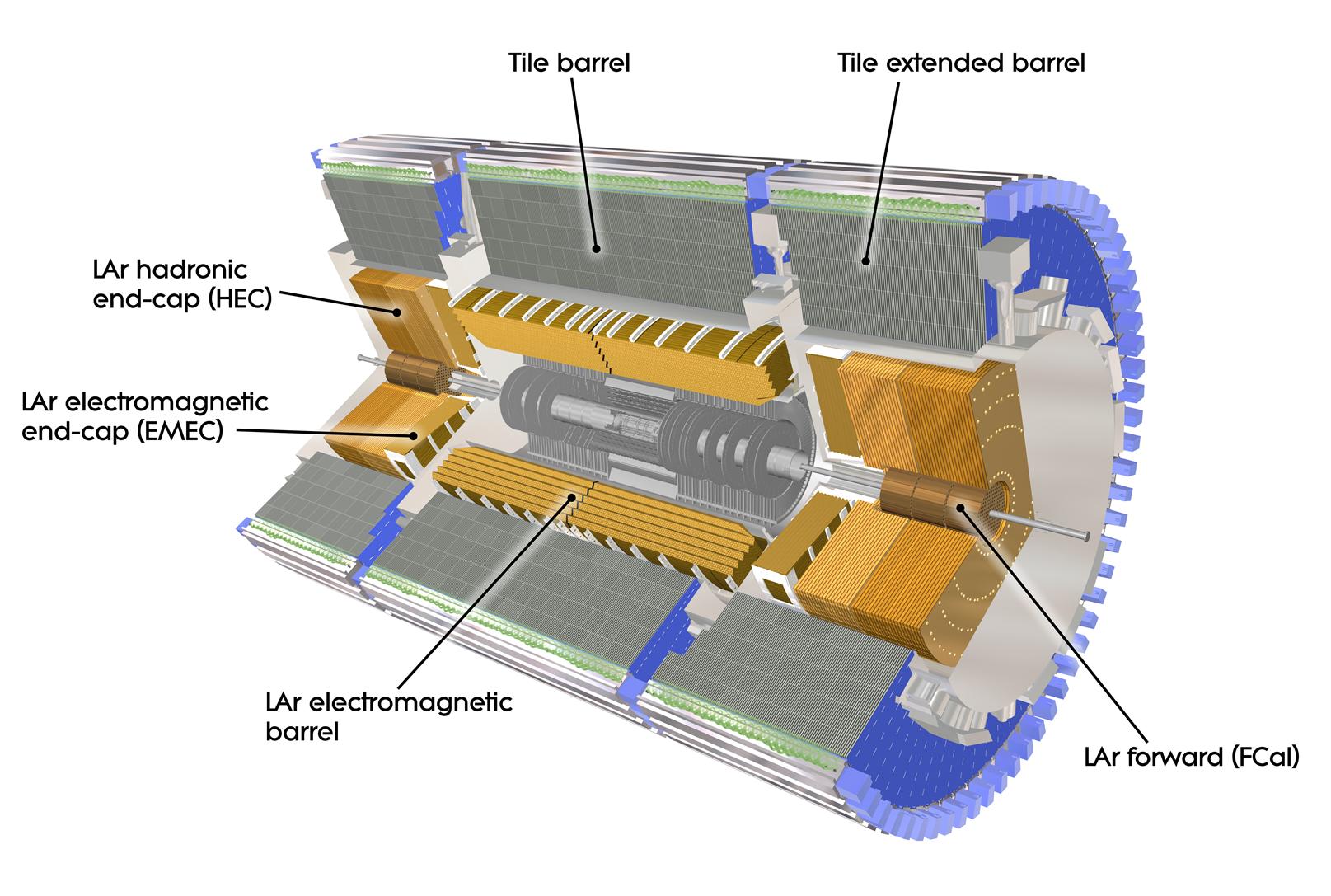}
\caption{Schematic view of the ATLAS calorimeter systems.}
\label{fig:ATLAScalo}
\end{figure}

A schematic view of the ATLAS calorimeter systems is shown in
Figure~\ref{fig:ATLAScalo}.
Precision electromagnetic (EM) calorimetry is provided by 
barrel (EMB) 
and endcap (EMEC) 
accordion geometry lead/liquid-argon (LAr) sampling calorimeters. 
The EMB shares the cryostat with the superconducting solenoid which
provides the magnetic field for the ATLAS inner tracker.
The EMEC calorimeters 
are contained in separate endcap cryostats,
together with the copper/LAr hadronic endcap (HEC) 
calorimeters providing hadronic coverage in the endcaps,
and forward copper-tungsten/LAr (FCal) calorimeters 
covering the very forward regions.
Surrounding the LAr calorimeters is the steel/scintillator-based
Tile hadronic calorimeter, which is divided into a central
Tile barrel and separate Tile extended barrel calorimeters in
the more forward regions.

The detectors of both the LAr and the Tile calorimeter systems are sufficiently robust
to withstand the difficult conditions of the HL-LHC phase, and
sufficiently radiation hard to tolerate the total doses expected by
the end of the HL-LHC operation period. As a result, no changes to the calorimeter detectors are going to be made, and the ATLAS HL-LHC
calorimeter upgrades focus only on electronics.

The upgrade of the readout electronics for both the LAr and
Tile calorimeters is necessary to meet 
the ATLAS physics goals in the demanding HL-LHC conditions,
with up to 200 $pp$ collisions per
bunch crossing at the bunch crossing rate of 40 MHz.
Triggering at the HL-LHC is very challenging, particularly since it
is critical to keep trigger energy thresholds low in order to 
maintain high efficiency for important physics processes (eg. Higgs
boson production) at the electroweak scale. 
The original electronics were designed with on-detector
pipelines serving as buffers of the calorimeter signals during
the first-level (L1) trigger latency of up 2.5~$\mu$s, and 
to digitize and read out the full granularity
at a L1 trigger rate up to a maximum of 100~kHz. With the
increased pileup and corresponding backgrounds 
at the HL-LHC, maintaining these limitations
would require unacceptably high trigger thresholds. In
addition, there are significant concerns
about whether the current on-detector
frontend (FE) electronics could survive the HL-LHC era,
due to either the increased radiation levels to be integrated at the HL-LHC, or to just its longevity in general, given the
current electronics were produced and installed before the
start of Run 1 at the LHC in 2010. To overcome these problems,
new readout electronics are being developed for the HL-LHC phase
for both the LAr and Tile calorimeters, designed with 
a free-running architecture,
where all channels are digitized at the 40~MHz 
bunch crossing rate, and all the digitized data are 
immediately transmitted off-detector. This scheme eliminates the
on-detector pipelines and allows the calorimeter
data with full 
granularity and full precision to be used in the high-level 
trigger algorithms. 

\subsubsection{ATLAS Liquid Argon Calorimeter Upgrade}


The ATLAS LAr calorimeter electronics must read out the  
182,468 active calorimeter channels. The performance requirements
are dominated by the ECAL specifications, which include
a 16-bit dynamic range (with energies in a single channel ranging from a noise level of about 50~MeV up to
a maximum of about 3~TeV), nonlinearity below 0.1\% in the 
electroweak energy range, and
an energy resolution of $\sigma_E / E = 10\% / \sqrt{E} \oplus 0.7\%$ (with $E$ measured in GeV and $\oplus$ denoting
addition in quadrature). The LAr calorimeter electronics upgrade is described 
in detail in Ref.~\cite{atlasLARTDR}.

\begin{figure}[h]\centering
\includegraphics[width=\textwidth]{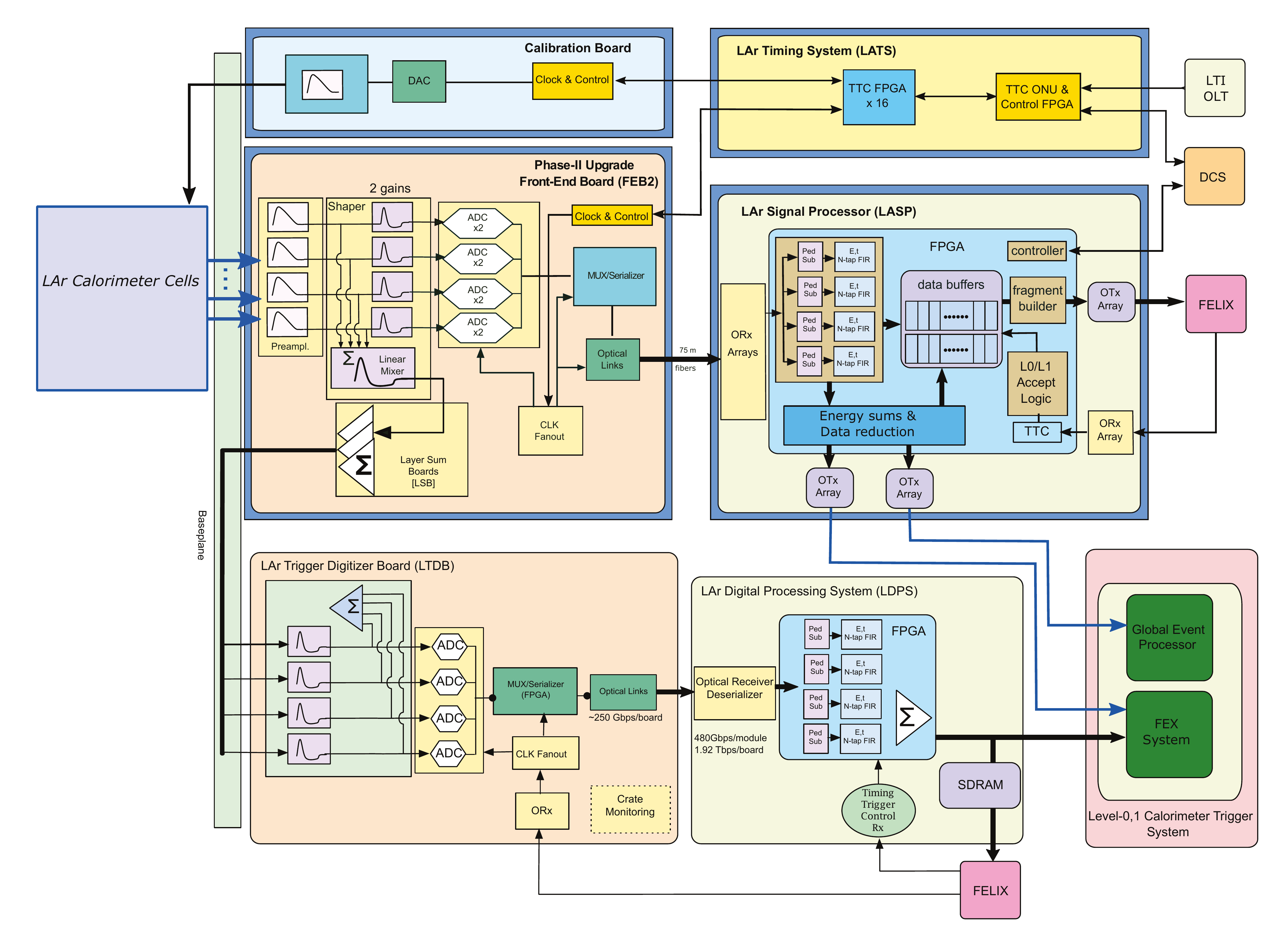}
\caption{Schematic block diagram of the HL-LHC LAr calorimeter readout architecture.
}
\label{fig:LAr:arch}
\end{figure}

A block diagram of the HL-LHC LAr readout 
is shown in Figure~\ref{fig:LAr:arch}.
The main elements to be developed 
are shown in the upper part of the figure, and include the new on-detector ``FEB2'' front-end
boards and new off-detector ``LASP" (LAr Signal Processor) boards that will implement the
precision readout path. New on-detector Calibration (CALIB) boards, that inject precision
pulses into the calorimeters for calibration purposes, will also be
needed, as well as new off-detector ``LATS'' (LAr Timing System) boards for distributing 
clock and control signals to the FE electronics.
(The LAr Trigger Digitizer Boards (LTDB) and the LAr Digital
Processing System (LDPS), shown in the lower part of
the block diagram, have already been installed as
part of the Phase-I upgrade of the LAr trigger, 
and will remain operational in the HL-LHC phase.) 

The FEB2 boards receive the signals from the calorimeter cells and perform analog processing, 
including amplification, shaping and a split into two overlapping linear gain scales.
Both gain scales are digitized at 40~MSPS, and 
the digital signals are multiplexed, serialized,  and transmitted optically  off detector to
the LASP board. The LASP boards
receive the digitized data from the FEB2 boards
and apply digital filtering to the signals of each LAr calorimeter cell. This 
is accomplished by an FPGA and allows the determination 
of calibrated cell energies and signal times with respect to the bunch crossing time 
(including an active correction of out-of-time pileup). The data are then 
buffered until a trigger accept signal is received.
For triggered events, the data are sent to the data acquisition (DAQ) system
for further processing and offline storage.  Since the LASP boards will
be installed off-detector, they do not need to be radiation
tolerant and can take advantage of utilizing fast commercial digital
electronics components.

A key challenge in developing the on-detector FE electronics, including the new FEB2 and CALIB boards, is that
they must satisfy radiation-tolerance requirements which,
including various safety factors, are 
a total ionizing dose (TID) of 1.3~KGy, a total non-ionizing
energy loss (NIEL) equivalent to $4.3\times 10^{13}$~neutrons/cm$^2$,
and a total number of hadrons capable of producing single-event-effects (SEE) of $1.1\times 10^{13}$~h/cm$^2$.
These requirements limit the use of commercial-off-the-shelf
(COTS) components. Instead, a number of custom ASICs must
be developed. 

The key custom components used on the FEB2
include the ``ALFE'' preamplifier/shaper ASIC, 
the ``COLUTA'' 40~MSPS ADC, the ``lpGBT'' 10~Gbps serializer ASIC,
and a VCSEL driver ASIC included as
part of the ``VTRx+'' optical transceiver.  While the lpGBT and
VTRx+ are developed by CERN-based collaborations for general use
at the HL-LHC, the ALFE and COLUTA are being
developed specifically for LAr. With the exception of ALFE, the FEB2 
custom
ASICs are all implemented in a 65~nm CMOS process which has
been demonstrated to be radiation-tolerant well 
beyond the LAr HL-LHC requirements.
Other advantages of this process include its high speed and 
low power. However, the 1.2~V power rail for the process does 
pose challenges for achieving the analog
precision requirements of some applications, particularly for
the COLUTA ADC which requires at least 14-bit dynamic range.
The ALFE uses instead a radiation-tolerant 130~nm
CMOS process, allowing a 2.5~V power rail to be used for the
input stage of the preamplifier,  which must handle the full 16-bit dynamic range of the calorimeter signals,

In addition to also using the
lpGBT and VTRx+ for clock and control functions, the CALIB
board faces special requirements, namely the need for a
16-bit DAC to precisely adjust the calibration pulse
amplitude, and a high frequency switch capable of handling
voltages in the range 5 - 10~V. The DAC has been implemented
in the ``LADOC'' ASIC in 130~nm CMOS, while the ``CLAROC''
switch ASIC
is being implemented in a 180~nm CMOS HV SOI process.


\subsection{CMS High Granularity Endcap Calorimeter}
\label{subsec:CMS}

At the HL-LHC proton-proton (pp) collision rates will reach 8 GHz with a 40 MHz bunch crossing (BX) rate and 200 pp collisions per BX.  These extreme collision rates present major challenges at multiple stages in detector readout and trigger systems for calorimeters, which must provide high precision information over a wide dynamic range at the full 40 MHz bunch crossing rate.   

The High Granularity Endcap Calorimeter (HGCAL) is a major upgrade of the CMS detector planned to for the HL-LHC. The HGCAL is described in detail in Ref.~\cite{cmsTDR}. This ``imaging calorimeter,'' which includes over 600\,m$^2$ of active silicon and over 6 million readout channels, is composed of $50$ layers of active shower-sampling media interleaved with dense absorber. The active medium of the $28$-layer front electromagnetic compartment is silicon, while the $22$-layer rear hadronic compartment includes both silicon and plastic scintillator instrumented with silicon photomultipliers.  Silicon layers are tiled with $8$'' hexagonal sensor modules, with each module including either 432 or 192 sensor channels depending on its location in the detector.  For the 40 MHz trigger readout, these sensor channels are ganged into $48$ logical trigger cells (TC) per module.  Scintillator layers are tiled with 64-channel tileboards.

On-detector (front-end) electronics must operate with low latency and low power dissipation in a high radiation environment, necessitating the use of custom ASICs.  The HGCAL front-end (FE) system reads out detector sensors on two data paths.  The trigger (TRG) path transmitted reduced data at 40 MHz for use in the trigger system.  The data acquisition (DAQ) path reads out data at an average rate of about 750 kHz for triggered events.  The full CMS trigger system has a latency of 12.5 $\mu$s (500 BX).  On the TRG path, the FE electronics have a combined latency of about 36 BX.  The total power budget for all FE electronics is about 20 mW per channel.

HGCAL uses a multi-stage, multi-ASIC FE system including the HGCROC, ECON-T, ECON-D, and lpGBT ASICs providing major functionality. HGCROC receives signals from up to 72 sensor channels and provides low noise and high precision measurements of signal amplitude over a 16-bit dynamic range from 0.2 to 10000 fC and signal time of arrival (ToA) with approximately 50 ps precision per channel for signals exceeding about 10 fC.  For each TC and BX, HGCROC transmits a 7-bit floating point encoding of the amplitude. For each sensor channel, HGCROC buffers the 10-bit amplitude (Q), 10-bit ToA, and 10 bits amplitude from the signal in the previous BX (Q-1), which is used to correct for potential leakage of charge signal from one BX into the next, and transmits the full 30-bit data for all sensor channels for each triggered event.  

The ECON-T ASIC receives the 7-bit TC data from up to six HGCROCs (48 TC total) on up to 12 1.28 Gbps inputs and selects / compresses the data according to one of four user-selectable algorithms.  The "threshold" (TS) algorithm simply reads out TC with amplitude exceeding a programmable threshold.  The "best choice" (BC) algorithm sorts the TC according to amplitude and reads out the N TC with highest amplitude, where N is user programmable.  The "super TC" (STC) algorithm sums together TC from a fixed region and transmits both the summed amplitude and the address of the TC in the region with highest amplitude.  Finally, ECON-T includes a configurable neural network (NN) that the user can train to compress the $48 \times 7$-bit input into 16 outputs each with 3 bits.  The threshold algorithm is the baseline choice and provides excellent physics performance but brings the complexity of a variable latency and data format.  Similar physics performance can be achieved with fixed latency and data format by using a hybrid approach with the BC algorithm in the front layers of HGCAL (dominant location for energy deposition by electromagnetically interacting particle) and the STC algorithm in the back layers (dominant location for energy deposition by hadronically interacting particles).  The NN algorithm also provides similar physics performance and can be continually reconfigured to address changing detector or LHC beam conditions.  

For each triggered event, the ECON-D ASIC receives 32 bits of raw data (Q, Q-1, ToA, and two status bits) from each of the 72 sensor channels from each of up to 6 HGCROCs on up to 12 1.28 Gbps input channels. The ECON-D ASIC applies a threshold-based zero suppression algorithm to decide which channels to read out and for each channel whether to transmit Q-1 or ToA in addition to Q.  The ECON-D and ECON-T ASICs transmit their outputs on 1-12 1.28 Gbps links to 1-2 lpGBT ASICs, which serializer the data from multiple ECON ASICs into 10.24 Gbps stream which is transmitted off-detector via radiation tolerant VTRx$+$ modules.

Two off-detector, back-end (BE) electronics systems, based on FPGAs and the Serenity ATCA boards, receive the data from ECON-T on the TRG path and ECON-D on the DAQ path. The BE TRG system is a two stage system, in which Stage 1 performs calibration and arranges the data according to detector geometry, and Stage 2 performs three-dimensional clustering of TC into reconstructed particle showers for use by the central CMS TRG system. The latency budget for the BE TRG system is about 3.5 $\mu$s or 140 BX. The BE DAQ system receives data and builds events for transmission to the central CMS DAQ system.

\section{Calorimeter Readout at Future Colliders}
\label{sec:future}

\subsection{Dual Readout Calorimetry}
\label{subsec:dualreadout}

Dual readout calorimeters can achieve exceptional energy resolution by reading out two different but complementary signals that together provide more precise information about the shower. 
The benefits and challenges for dual-readout calorimetry for future collider experiments and the associated readout electronics are described in detail in Ref.~\cite{dualreadout_whitepaper}.  Like other calorimeter technologies, the physics demands for future dual readout calorimeters require high transverse granularity, some amount of longitudinal segmentation, charge readout with excellent precision and high dynamic range, and precise measurement of signal arrival times.  The unique readout challenge for a dual-readout calorimeter is simultaneous low-latency extraction of charge signals from both scintillation and Cherenkov light, which requires complex on-detector electronics and which doubles the data relative to a traditional single-signal detector.


\subsection{FCC-ee and linear electron colliders}

Proposed calorimeters for future $e^+e^-$ colliders span a wide variety of materials and technologies. 
These designs are motivated by physics specifications, namely to ensure that calorimeters can make precision measurements on the Higgs boson and other Standard Model (SM) particles, while providing sensitivity to beyond the SM searches. 


Several challenges and considerations pertain to the development of readout electronics for calorimeters in $e^+e^-$ machines.
Novel calorimeter designs such as particle flow and dual readout will require dedicated R\&D in the corresponding readout systems. 
Common to almost all new calorimeter proposals is a highly granular design, which requires a high density of readout channels and thus potentially high power consumption in a small detector area. 
The need for precise timing resolution, for example in the case of 4D calorimetry, also introduces a need for smaller feature size in ASICs and keeping pace with CMOS evolution.
Finally, the use of novel materials such as silicon and fibers introduce new challenges with cold readout or high sampling frequencies, for which there is less precedent on best practices from the LHC experiments. 
In this section, specific collider considerations and calorimeter proposals are discussed in the context of future readout system design.

\subsubsection{Readout for Linear Electron-Positron Colliders}

There are several well-developed proposals for linear electron-positron colliders, such as the International Linear Collider (ILC) or the Cool Copper Collider ($C^3$).
Several future calorimeter designs have been prepared to provide performant calorimetry in line with the physics requirements of these experiments. 
The most unique feature of these colliders from a readout perspective is the ``power pulsing" operation, wherein the beam crossing of the accelerator has a very low duty cycle (10$^{-3}$)~\cite{ILC}. 
This offers an opportunity to reduce heat load and power budget in the electronics by fast switching off during the breaks between bunch trains.
This can be especially critical in a dense calorimeter, where high channel multiplicity and small infrastructure spatial requirements mean that power consumption must be reduced to a minimum. 
Studies have been performed by the CALICE Collaboration to investigate dedicated ASIC designs that accommodate fast logic and low-noise power cycling during operation with strong spatial constraints~\cite{powerpulsing_CALICE, powerpulsing_CALICE_pflow}.

Two specific examples of calorimeter designs for future linear $e^+e^-$ colliders are the Silicon Detector (SiD) and the International Large Detector (ILD), both for the ILC. 

The SiD ECAL uses silicon as an active material, similar to that of the CMS HGCAL upgrade discussed in Section~\ref{subsec:CMS}, interleaved with tungsten absorbers.
It has excellent jet energy measurement due to a particle flow reconstruction technique that relies on fine longitudinal and transverse segmentation. 
Readout of the silicon pixels can be performed by a single chip, for example the KPiX ASIC, which is bump bonded to the sensor and reads time-over-threshold with a 2000 minimum ionizing particle (MIP) dynamic range.
The HCAL for SiD relies on scintillators read out by silicon photo-multipliers (SiPMs), which can allow for electronics to be included on-chip, providing lower operating voltages and higher speed than traditional photo-multiplier tubes.

The ILD ECAL also employs a highly granular particle flow calorimeter, but implemented via sampling of silicon diodes, or alternately through scintillator strips. 
Two options are also studied for the HCAL, based on either SiPM-on-tile technology for scintillator readout, or resistive plate chambers.
In both calorimeters, issues with signal-to-noise ratio and ghost hits in SiPM readout motivate the study of double SiPM readout for longer strips, which can maintain the stringent timing and spatial resolution requirements~\cite{ILD_SiPM}.
Prototype chip designs have been developed in the ILC context for SiPM readout~\cite{SPIROC}.


\subsubsection{Readout for Circular Electron-Positron Colliders} 

As with the ILC, future circular $e^+e^-$ colliders such as FCCee have several calorimeter proposals.
In the case of FCCee, there are two main options, the ``CLIC-Like Detector" (CLD)~\cite{CLD} and the Innovative Detector for Electron-positron Accelerator (IDEA)~\cite{IDEA} detector. 
The detector requirements are very similar to those of the ILC and the calorimeter specifications follow. 

Like all $e^+e^-$ calorimeters discussed so far, CLD uses particle flow calorimetry with the aim of reconstructing visible particle four-vectors with high precision.
It also shares much of the SiD design, in that the ECAL has a silicon-tungsten sandwich structure, and the HCAL has steel absorbers interleaved with scintillator tiles that rely on SiPM readout. The IDEA detector design differs significantly by introducing a novel copper-based dual-readout fibers calorimeter, where the elimination of longitudinal segmentation allows for finer lateral segmentation with the same number of electronic readout channels.
The fibers are read out by SiPMs, and the readout design in this scenario is driven by the challenges discussed in Section~\ref{subsec:dualreadout}.





\subsection{FCC-hh }

A long term prospect for a future proton-proton collider (FCC-hh) is in the early planning stages. 
With the goal of opening a new energy regime for discoveries, the FCC-hh  would deliver collisions at energies up to 100~TeV with instantaneous luminosities up to $3 \times 10^35$~cm$^{-2}$s$^{-1}$. However, searching for new physics decaying to final states of photons, electrons, muons, taus or jets will place unprecedented requirements on the spatial and energy resolution of the detectors. The spatial resolution requirements are driven by the need to resolve the narrowly collimated decay products, for example, photon showers from highly boosted particles. Meanwhile, the energy resolution requirement arises from the desire to measure the energy deposits of minimum ionising particles up to multi-TeV energy electrons and photons. Therefore, this challenging environment provides an opportunity to push calorimeter readout technology as far as possible. 

\begin{figure}[ht]
\begin{subfigure}{0.8\textwidth}
\includegraphics[width=1.0\linewidth]{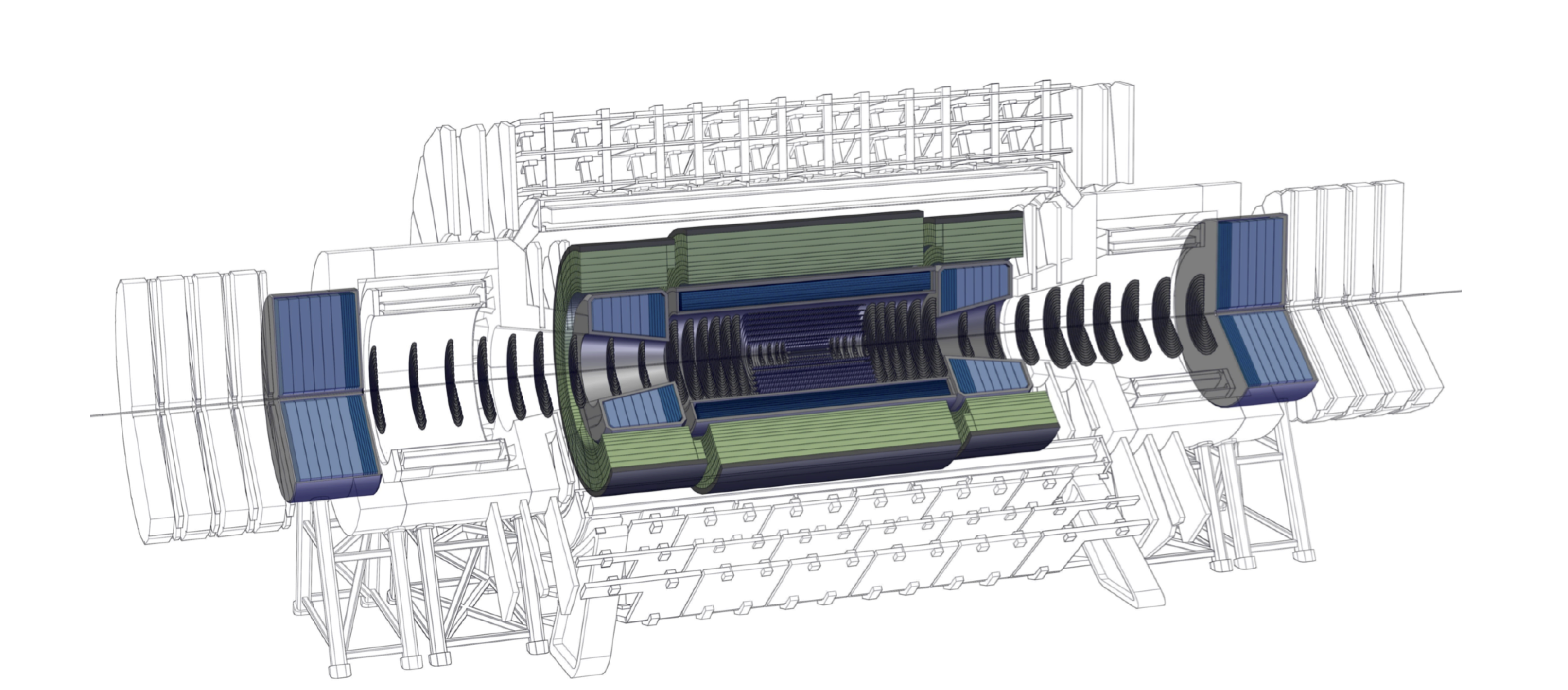} 
\end{subfigure}

\caption{Reference FCC-hh calorimeter (from~\cite{FCC-hh}).} 
\label{fig:fcc-hh}

\end{figure}

The potential reference design for the FCC-hh calorimeters is shown in Fig.~\ref{fig:fcc-hh}. The cylindrical design, at 50~m long and 20~m in diameter, is similar in size to the ATLAS detector but in mass to the CMS detector. It consists of a tracker cavity surrounded by an ECAL with a thickness of 30 $X_0$ and an HCAL with a thickness of 10.5 interaction lengths. The ECAL is based on LAr for its well understood radiation hardness, while the the HCAL is based on scintillating tiles with lead absorbers. 

\begin{figure}[h]

\begin{subfigure}{0.8\textwidth}
\includegraphics[width=1.0\linewidth]{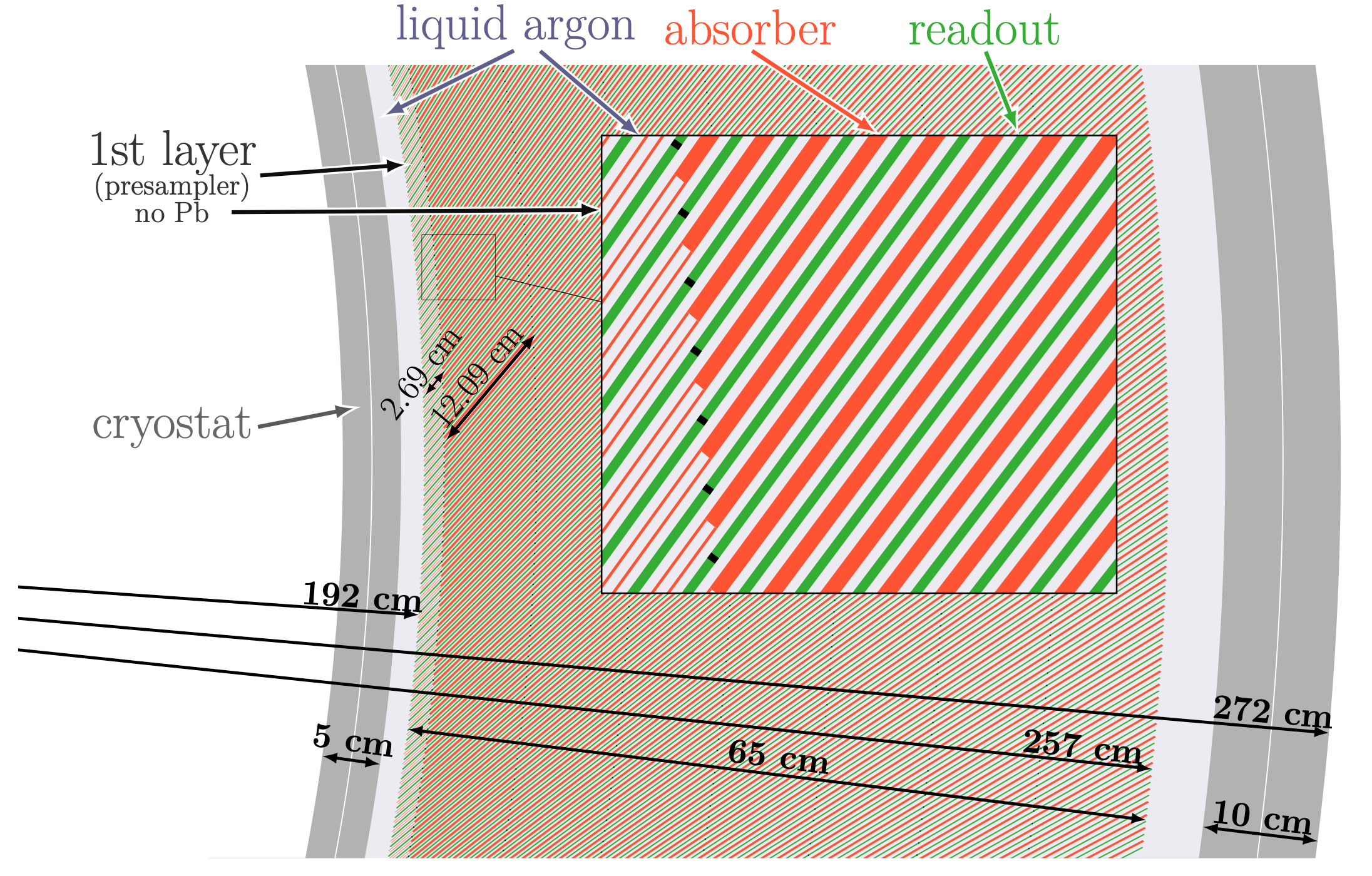} 
\end{subfigure}

\caption{The cross section of the electromagnetic barrel calorimeter showing the extremely fine segmentation of the design. From~\cite{FCC-hh}.} 
\label{fig:caloXS}
\end{figure}

The ECAL calorimeter system for this detector (described in~\cite{FCC-hh})  is composed of sampling calorimeters with liquid Argon as the sampling medium. The high granularity and fine segmentation of the design, resulting in over two million channels, is indicated in Fig.~\ref{fig:caloXS}, and is similar to the CMS HGCAL. The high granularity will help facilitate the use of the particle flow technique 
combining measurements from several detectors to improve the final 4-momentum measurement of a particle. High granularity calorimetry would also enable improved vertex finding for neutral particles such as highly boosted photons. 

The ECAL has in some respects the most challenging requirements of any readout considered. Foremost, at a 100~TeV FCC-hh machine, the calorimeter must be able to provide accurate photon and lepton energy measurements from the electroweak scale ($\approx$~GeV) and up to energies of tens of TeV. This suggests a dynamic range of 16~bits, achieved through a multi-gain architecture. Given the progress in ``4D'' particle reconstruction and searches for long-lived particles, the desired time resolution is on the order of 30~ps. The dose near the location of the readout electronics is potentially 1-10~MRad (outside the LAr cryostat), with additional safety factors to be determined. The goal is to read out the over three million channels for the ECAL every 25~ns. 

The accomplishment of this readout will build on the electronics developed in both the CMS and ATLAS detector upgrades for the HL-LHC. However, significant R\&D is required to read out three million channels with high precision and dynamic range within a space and power requirements that are not likely to be significantly larger than today. This will necessitate exploring new readout architectures. For example, simply extrapolating from current work is infeasible, as higher channel density ASICs on more densely populated readout boards would need to accommodate an order of magnitude increase in channel readout, while extending the dynamic range. New technology nodes, combined with System-on-chip architectures, may ameliorate some of these challenges.

\section{Opportunities for Advancing Readout for Calorimetry}
\label{sec:innov}
%
%

\subsection{Taking advantage of infrastructure developments}

As discussed, the next generation of particle colliders will require challenging high-speed, high-precision electronic readout of high granularity (high-channel density) particle detectors. There are many challenges facing the design of the ASICs needed for the readout of these calorimeters. Fortunately, there are emerging developments in industry that may mitigate some of the difficulties. One example can be found in foundries  that offer a commercially available fully open hardware design and fabrication environment.

There are several advantages to using an open ecosystem. First, the lower costs (compared to traditional foundries) and faster turnaround for fabrication could significantly increase the pace of development. Instead of an annual submission cycle, a faster submission cycle in which a chip is being tested, while the next version is being fabricated, and the subsequent version is in design could be possible. Perhaps more important, an open ecosystem would improve the ease of collaboration with groups at universities, national laboratories, and international partners. Calorimetry readout design efforts could benefit from initial assessment of open source hardware developments.

\section{Summary and Conclusion }

Calorimeter will remain an important tool for future particle detectors. The unique requirement of high dynamic range, high precision, and high readout rates for signal amplitudes, combined with increasing granularity and precision timing present significant conceptual challenges. This places greater emphasis on the engineering challenges of delivering low noise, low power, radiation tolerant devices. However, the ongoing upgrades of the CMS and ATLAS detector provide a strategy for overcoming these obstacles. Carefully planning the future calorimeter readout systems and exploiting new technologies will allow the designers to deliver systems at future colliders that meet the stringent specifications. 










\end{document}